\documentclass[12pt,a4paper,fleqn]{article}

\usepackage[DIV12]{typearea}
\usepackage{amsmath,amsfonts,amssymb,mathrsfs}
\usepackage{graphicx}
\usepackage[footnotesize]{caption}
\usepackage{cite}
\usepackage{subfigure}
\usepackage{color}

\newcommand{\sltau}[1][]{{\widetilde\tau_{#1}}}

\newcommand{\brac}{\ensuremath{\phantom{\frac{1}{1}}\!\!\!\!}}
\usepackage{feynmp}

\newcommand{\stau}{{\widetilde\tau}}
\newcommand{\snu}{{\widetilde\nu}}

\newcommand{\eVdist}{\kern-0.06em}

\newcommand{\gev}{\:\text{Ge\eVdist V}}
\newcommand{\tev}{\:\text{Te\eVdist V}}

\newcommand{\CenterEps}[2][1]{\ensuremath{\vcenter{\hbox{\includegraphics[scale=#1]{#2.eps}}}}} 
\newcommand{\D}{\mathrm{d}}

\newcommand{\U}[1]{\ensuremath{\mathrm{U}(#1)}}

\allowdisplaybreaks

\allowdisplaybreaks[1]

\begin{document}

\begin{titlepage}

\begin{flushright}
TUM-HEP 695/08
\end{flushright}

\vspace*{1.0cm}

\begin{center}
{\Large\bf 
A note on the primordial abundance of stau NLSPs
}

\vspace{1cm}

\textbf{
Michael Ratz\footnote[1]{Email: \texttt{mratz@ph.tum.de}},
Kai Schmidt-Hoberg\footnote[2]{Email: \texttt{kschmidt@ph.tum.de}}
and Martin Wolfgang Winkler\footnote[3]{Email: \texttt{mwinkler@ph.tum.de}}
}
\\[5mm]

Physik-Department T30, Technische Universit\"at M\"unchen, \\
James-Franck-Stra\ss e, 85748 Garching, Germany

\end{center}

\vspace{1cm}

\begin{abstract}
\noindent 
In scenarios with a gravitino LSP, there exist strong BBN constraints on the
abundance of a possible stau NLSP. We find that in settings with substantial
left-right mixing of the stau mass eigenstates these constraints can be evaded
even for very long-lived staus. 
\end{abstract}

\end{titlepage}

\newpage

\section{Introduction}

Arguably, supersymmetry (SUSY) is one of the most plausible extensions of the
standard model (SM). Apart from its theoretical appeal, SUSY has the virtue of
providing a compelling dark matter candidate, the lightest supersymmetric
particle (LSP), which is stable if $R$ parity is conserved. A particular
interesting dark matter candidate is the gravitino, which evades direct
detection because all its interactions are suppressed by the Planck scale. The
hypothesis of a gravitino LSP may nevertheless be tested at the LHC if certain
conditions are met. First, the next-to-lightest
supersymmetric particle (NLSP) has to be charged, and second the gravitino mass
may not be too small. In this case, one could observe long-lived charged
particles in whose decays one could probe the properties of the LSP
\cite{Buchmuller:2004rq,Buchmuller:2004tm,Feng:2004mt,Brandenburg:2005he}. 
The collider phenomenology of such
scenarios has been explored in various studies 
\cite{Buchmuller:2004rq,Buchmuller:2004tm,Feng:2004mt,Hamaguchi:2004df,Feng:2004yi,Brandenburg:2005he,Martyn:2006as,%
Ellis:2006vu,Kitano:2008sa}.

There are several theoretical reasons which make it appear desirable to have
gravitino masses $m_{3/2}$ not
much smaller than the masses of the SM superpartners.
For instance, simple explanations of the  $\mu$ and $B\,\mu$ terms
seem to require not too small $m_{3/2}$ \cite{Kim:1983dt,Giudice:1988yz}.
Further, many simple mechanisms of baryogenesis, in particular leptogenesis
\cite{Fukugita:1986hr},
need rather high reheating temperatures $T_R$ \cite{Davidson:2002qv} which can be achieved
for gravitino masses of about $10\dots100\,$GeV
\cite{Bolz:2000fu,Pradler:2006hh} (i.e.\ at least the constraints from gravitino
overproduction can be satisfied).

However, there are severe constraints on such scenarios coming from cosmology. 
The observed primordial abundances of light elements produced in big bang
nucleosynthesis (BBN) allow to place stringent constraints on the number density
of long-lived particles whose decays happen during or after BBN  and induce
nuclear reactions that change the element abundances
\cite{Falomkin:1984eu,Khlopov:1984pf,Ellis:1984eq}.  While a neutralino NLSP is
strongly disfavored for gravitino masses in the \gev\ 
range~\cite{Feng:2004mt,Cerdeno:2005eu},  scenarios with a sneutrino NLSP are
essentially unconstrained but very hard to test experimentally
\cite{Kanzaki:2006hm,Covi:2007xj}. This makes a charged slepton, specifically a
stau, particularly
appealing as an NLSP candidate. The stau NLSP abundance and lifetime can
satisfy the limits obtained from BBN by considering NLSP decays alone
[\citen{Asaka:2000zh},\citen{Fujii:2003nr,Feng:2004mt,Ellis:2003dn,Cerdeno:2005eu,Jedamzik:2005dh,%
Steffen:2006hw,Buchmuller:2006nx,Cyburt:2006uv,Pradler:2006hh}]. However, as
pointed out in \cite{Pospelov:2006sc}, charged NLSPs form bound states with
light nuclei, which leads to a drastic overproduction of $^6$Li. This process, known as  
Catalyzed BBN (CBBN),
leads to strong constraints on the stau relic abundance, unless the NLSP lifetime 
is shorter than a few thousand seconds.

Several ways to circumvent BBN constraints have been discussed in the
literature. For instance, entropy production between NLSP freeze-out and the
start of BBN can dilute the NLSP abundance sufficiently to satisfy all
constraints even for long lifetimes
\cite{Buchmuller:2006tt,Pradler:2006hh,Hamaguchi:2007mp}. However,
in order to arrive at such scenarios one usually relies on new sectors which are
typically very hard to access experimentally. Alternatively, the NLSP can be
sufficiently short lived if the gravitino is very light, $R$ parity is slightly broken
\cite{Takayama:2000uz,Buchmuller:2007ui} or the superpartner mass spectrum is
sufficiently heavy \cite{Kersten:2007ab}. However in these cases it is practically
impossible to test the nature of the LSP.

The purpose of this study is to point out that there are regions within the 
parameter space of the minimal supersymmetric extension of the standard model 
(MSSM) where the relic stau abundance is strongly suppressed such that the
bounds from CBBN can be evaded even for long stau lifetimes.  As
we shall see, small thermal relic abundances of staus occur in parameter regions
with a substantial left-right mixing of the stau mass eigenstates, where the
annihilation into Higgs bosons is greatly enhanced.

The paper is organized as follows. In the next section we will introduce the stau-Higgs coupling
and calculate the Higgs channel cross section. 
Section~\ref{sec:muconstraints} is devoted to a discussion
of theoretical constraints on trilinear couplings between Higgs and $\stau$
fields.
In Section~\ref{primordial}  we review the relevant BBN constraints and discuss
the stau relic abundance. Continuing with Section~\ref{sec:scenarios} we
introduce three scenarios within the MSSM in which a strong suppression of the
stau relic abundance can be achieved such that all cosmological constraints can
be evaded. Finally in Section~\ref{sec:Prospects} we briefly discuss the
implications of our scenario for the LHC.

\section{Annihilation into Higgs bosons}

In the early universe, superpartners are copiously produced; usually they are
assumed to be in thermal equilibrium. As the universe cools down, they will
cascade into staus, which we assume to be the NLSPs. Since staus are metastable,
until the BBN era their abundance will only decrease due to annihilation.
In most analyses performed so far, the lightest stau is assumed to be purely
right-handed. Then, for its freeze-out, only electroweak annihilation processes have to
be considered. The couplings governing the relevant reactions are either the
electric charge $e$ or the $\U1_Y$-coupling $g_Y=e / \cos{\theta_\mathrm{W}}$,
where $\theta_\mathrm{W}$ denotes the Weinberg angle. These couplings are rather
small, leading to a relatively large stau abundance after freeze-out
\cite{Asaka:2000zh},
\begin{equation}
 Y_{\stau_\mathrm{R}}~\gtrsim~10^{-13} \quad  \text{for} \quad {m_{\stau_1}} \gtrsim 100\gev\;,
\end{equation}
where the abundance  $Y\equiv n/s$ is defined as the ratio of number and entropy densities. 
Such a large relic stau abundance is allowed by CBBN only if the gravitino is
very light and the stau lifetime accordingly short. If the lighter stau has a
left-handed component, the electroweak annihilation cross section gets enhanced
due to its $\mathrm{SU}(2)_\mathrm{L}$ couplings, but as CBBN bounds are very
tight, the inclusion of further gauge interactions changes the situation only
marginally. On the other hand, we shall see that in the case of substantial
left-right mixing in the stau sector, the couplings between staus and Higgs
bosons can get significantly enhanced, thus greatly suppressing the stau relic
abundance.

\subsection{Coupling of staus to Higgs bosons}

To find the regions of parameter space where this annihilation reaction is
important, we now turn to the Lagrangean term which describes the couplings
between the light stau and the light Higgs. In our analytic discussion, we
make a couple of simplifying assumptions; later, in Section~\ref{sec:scenarios},
we will take into account all interactions and states. 

We shall assume that there is no generation mixing in the slepton sector which
is suggested by flavor constraints. Furthermore we take $\mu$ and $A^\tau$ to be
real parameters. Then  the relevant terms read\footnote{There exist different
sign conventions for the A-parameter.  Here we follow \cite{Drees:2004jm}.}
\begin{eqnarray}
 \mathscr{L}_{\sltau[1] \sltau[1] h}&=&
 \frac{g_2}{2M_W} 
 \left\lbrace \brac M_W^2\, \sin{(\alpha +\beta)}\, 
 \left[ \left( \tan^2{\theta_\mathrm{W}}-1\right) \,
 \cos^2{\theta_{\sltau}} -2 \tan^2{\theta_\mathrm{W}}\, 
 \sin^2{\theta_{\sltau}} \right] \right. \nonumber\\
 &&\left.\phantom{\frac{g_2}{2M_W}\lbrace }
 {}+m_\tau\, \frac{\mu\, \cos\alpha- A^\tau\, \sin\alpha}{\cos\beta} 
 \sin{2\theta_{\sltau}} 
 + 2m_\tau^2\, \frac{\sin\alpha}{\cos\beta} \right\rbrace \; h 
 \,\sltau[1]^+\sltau[1]^- \;.
\end{eqnarray}
For simplicity we assume that the Higgs bosons except $h$ be relatively heavy
($\gtrsim 300 \gev$), which is the case for all models we are considering later.
This allows us to work in the `decoupling limit' where the mixing parameter
$\alpha$ can be written as $\alpha \simeq \beta-\pi/2$ and therefore $\cos\alpha
\simeq \sin\beta$ and $\sin\alpha \simeq -\cos\beta$.  The leading term of the
Lagrangean which couples the lightest stau to the Higgs is then given by
\begin{eqnarray}
 \mathscr{L}_{\sltau[1] \sltau[1] h}&=&
 \frac{g_2}{2M_W}\,
 \sin{2\theta_{\sltau}}\; m_\tau\,  \left\lbrace \mu\, \tan\beta+ A^\tau 
 \right\rbrace \; h \,\sltau[1]^+\sltau[1]^- 
\nonumber \\
&=&
-\frac{g_2}{2M_W}\,
\sin{2\theta_{\sltau}} \; m_{\stau_{\mathrm{LR}}}^2\; h
\,\sltau[1]^+\sltau[1]^-\;.\label{eq:StauHiggsCoupling}
\end{eqnarray}
Here $m_{\stau_{\mathrm{LR}}}^2$ denotes the off-diagonal
element of the $2\times2$ stau mass matrix (cf.\
Appendix~\ref{app:StauMassMatrix}).

\subsection{Higgs channel cross section}

\begin{figure}[h]
\centerline{\subfigure[{}]{\CenterEps{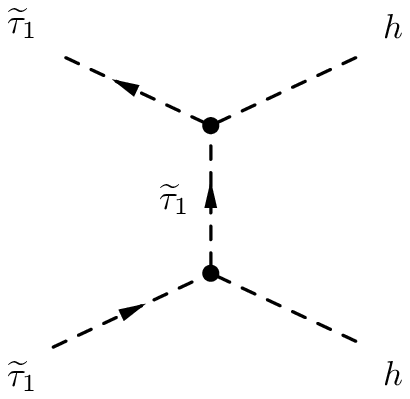}}\quad\quad\subfigure[{}]{\CenterEps{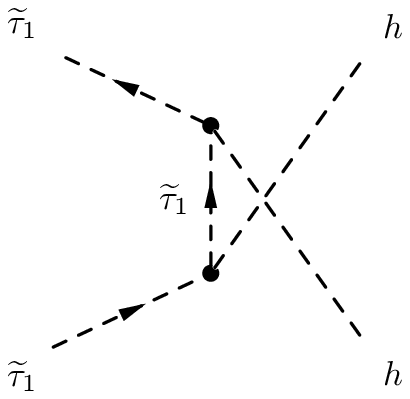}}}
\caption{Stau annihilation into Higgs bosons.}
\label{fig:StauHiggsAnnihilation}
\end{figure}

With the stau-Higgs coupling \eqref{eq:StauHiggsCoupling}, we can now calculate
the cross section for the annihilation of the light staus into light Higgs
bosons,
\begin{equation}\label{eq:StauHiggsAnnihilation}
 \stau_1^{+}+ \stau_1^-~\rightarrow ~h+h\;.
\end{equation}
The contributing Feynman diagrams are shown in
Figure~\ref{fig:StauHiggsAnnihilation}.
In our approximation, we consider only the exchange of $\stau_1$, neglecting the
exchange of $\stau_2$. 
In zeroth order of a velocity expansion, the thermally averaged annihilation cross section 
$\langle\sigma_\mathrm{ann}\,v\rangle$ is equal to the cross section $\sigma_\mathrm{ann}$ 
times the relative velocity $v_\mathrm{rel}$ of the incoming staus. We obtain
\begin{equation}\label{eq:ThermalCrossSection}
 \langle\sigma_\mathrm{ann}\,v\rangle~\simeq~\sigma_\mathrm{ann}\,v_\mathrm{rel}
 ~=~
 \frac{1}{16 \pi}\left( \frac{g_2}{2M_W}\,
 \sin{2\theta_{\sltau}}\; m_{\stau_{\mathrm{LR}}}^2\right)^4
 \frac{\sqrt{m_{\stau_1}^2-m_h^2}}{m_{\stau_1}^3(2 m_{\stau_1}^2-m_h^2)^2} 
\;.
\end{equation}
Clearly, this annihilation cross section becomes important for sizable left-right mixing
and relatively small stau masses. 

\subsection{Comparison with electromagnetic cross section}

It is instructive to compare the cross section \eqref{eq:ThermalCrossSection}
with a typical electroweak cross section. For example the annihilation cross
section of staus into photons is given by \cite{Asaka:2000zh} $\langle
{\sigma_\mathrm{ann}}_{\;\sltau^+\,\sltau^-\rightarrow \gamma\,\gamma}\,v\rangle
\simeq 4 \pi \alpha^2/m_{\stau_1}^2$. 
If $m_{\stau_1}$ and $m_h$ are not too close, one has
\begin{equation}\label{eq:RatioAnnihilationCrossSections}
 \frac{\langle {\sigma_\mathrm{ann}}_{\;\stau^+\,\stau^-\rightarrow
 h\,h}\,v\rangle}{%
 \langle {\sigma_\mathrm{ann}}_{\;\stau^+\,\sltau^-\rightarrow \gamma\,\gamma}\,v\rangle}
 ~ \sim~
 \left( \frac{\tan\beta}{50}\right) ^4\,
 \left( \frac{\mu_\mathrm{eff}}{2 m_{\stau_1}}\right) ^4\;,
\end{equation}
where $\mu_\mathrm{eff}=\mu\, \sin{2 \theta_{\sltau}}$.
Hence, for $\mu_\mathrm{eff}>2m_{\stau_1}$ the annihilation cross section is dominated by the
Higgs channel. Even for an order one ratio $\mu_\mathrm{eff}/2m_{\stau_1}$ one obtains a
dramatic reduction of the primordial stau abundance.
However, as we shall discuss next, there are constraints on the ratio
$\mu_\mathrm{eff}/2m_{\stau_1}$, implying that the reduction cannot be
arbitrarily strong.

\section{Theoretical constraints on $\boldsymbol{\mu}$}

\label{sec:muconstraints}

The enhancement of the stau annihilation cross section relies on a large
trilinear Higgs-stau coupling, which might lead to an unwanted (color and)
charge breaking (CCB) minimum of the potential. Such minima might be acceptable
if the `physical'  vacuum is sufficiently long-lived. In what follows, we will
first analyze the tree-level potential and see that in the interesting regions
of parameter space there is indeed an unphysical minimum. We will discuss
tunneling to this vacuum. Then we will discuss quantum corrections to the
potential and see that they lift the unwanted minimum and possibly even make it
disappear. Finally, we will comment on constraints on $\mu$ arising from
unitarity.

\subsection{CCB constraints at tree level}

The relevant field space is given by two real fields,
$\stau=\frac{1}{\sqrt{2}}\text{Re}(\stau_1)$ and $h$. 
The corresponding scalar potential  around the electroweak
vacuum can be written as
\begin{equation}
 V~=~\frac{1}{2}m_{\stau_1}^2\,\stau^2+\frac{1}{2}m_h^2\,h^2
 +b_h\,h^3+b_{\stau\stau h}\,h\,\stau^2
 +\lambda_h\,h^4+\lambda_{\stau}\,\stau^4+\lambda_{\stau h}\,\stau^2\,h^2\;,
\end{equation}
where at tree level $m_h \simeq 90 \gev$, $b_h\simeq 17\gev$, $\lambda_h\simeq
\lambda_{\stau h}\simeq 0.018$, $\lambda_{\stau}\simeq 0.028$ and the trilinear
coupling $b_{h\stau\stau} = - \frac{y_\tau}{\sqrt{8}}\mu_\mathrm{eff}$. For very
negative $b_{h\stau\stau}$ values there is a second minimum of the tree-level
potential. We have searched for minimal paths connecting the electroweak vacuum
with the second, deeper minimum. It turns out that the relevant field direction
is always very close to $x=\frac{1}{\sqrt{2}}(\stau+h)$. The corresponding potential
along that direction can be written as
\begin{equation}
\label{pot}
V~=~ 
\frac{1}{4}(m_{\stau_1}^2+m_h^2)\,x^2 + 
\left(\frac{b_h}{\sqrt{8}}- y_{\tau} \frac{\mu_{\text{eff}}}{8}\right)\,x^3 
+\frac{1}{4} (\lambda_h + \lambda_{\stau} + \lambda_{h\stau})\, x^4 \;.
\end{equation}
In order to check whether the lifetime of the local minimum at $x=0$  
exceeds the age of the universe we have to calculate the so called bounce action $S_\text{B}$
along the lines of \cite{Coleman:1977py}, which should satisfy $S_\text{B} \ge 400$ \cite{Kusenko:1996jn}.
Using the potential \eqref{pot} with the coefficients given by their tree-level values
we find an upper bound
on the coefficient of the trilinear term
$|\frac{b_h}{\sqrt{8}}-\frac{\mu_\mathrm{eff}}{16}| \lesssim 33.5 \,\mathrm{GeV}$,
which translates into 
\begin{equation}
 \mu_\mathrm{eff} ~\lesssim~630\,\mathrm{GeV}
 \quad\text{(for $\tan\beta=50$ and $m_{\stau_1}=120\gev$)}\;. 
\end{equation}
Note that for  $\mu_\mathrm{eff} \lesssim 430\gev$ the tree-level potential 
does not exhibit a second, deeper minimum at all.

\subsection{Quantum corrections to CCB constraints}
\label{sec:QuantumCorrections}

It is well known that quantum corrections can change the tree-level picture drastically.
In order to analyze the situation properly, one has to take into
account radiative corrections to the potential.
This is a very complicated issue, and we refrain from performing a complete
analysis here. 
In order to get a feeling for the impact of radiative corrections
we include the standard stop loop correction (cf.\ e.g.\
\cite[p.~245~f.]{Drees:2004jm}).
The resulting effective potential is significantly steeper in the Higgs
direction, such that the physical Higgs mass can
violate the tree-level bound $m_h\le m_Z$. 
It also turns out that the cubic and
quartic coefficients in the Higgs potential get enhanced by $\sim 70\,\%$ and
$\sim 40\,\%$, respectively. This has important implications for the bounce action:
plugging the loop corrected values for these parameters into \eqref{pot} we find that
the metastable minimum has a sufficiently large lifetime for
\begin{equation}\label{eq:mubound2}
 \mu_\mathrm{eff}~\lesssim~780\,\mathrm{GeV}
 \quad\text{(for $\tan\beta=50$ and $m_{\stau_1}=120\gev$)}\;. 
\end{equation}
The second, deeper minimum exists only if $\mu_\mathrm{eff} \gtrsim 500 \,\mathrm{GeV}$.

There are similar effects, in particular for the potential in $\stau$ direction.
This issue has been studied in~\cite{Ferreira:2001tk}, where it was found that
there are no charge breaking minima at all after quantum corrections are taken
into account. 
Whether or not these statements also apply to parameter regions with large
$\tan\beta$ and $\mu_\mathrm{eff}$ will be studied elsewhere.

\subsection{Unitarity bound}

Further constraints on $\mu_\mathrm{eff}$ come from unitarity.
The unitary cross section for scalar particles can be calculated using partial
wave expansion. In the case of a non-elastic process it takes the form
\cite{Griest:1989wd}
\begin{equation}
 \sigma_\mathrm{unit}~=~\frac{4 \pi(2J+1)}{|\vec{p}_\mathrm{in}|^2}\;,
\end{equation}
where $J$ is the angular momentum of the partial wave and
$\vec{p}_\mathrm{in}$ is the three-momentum of one incoming particle.
The dominant contribution to the stau annihilation into Higgs bosons comes from $s$-wave scattering. 
Therefore the perturbative unitarity constraint relevant to our discussion is
\begin{equation}\label{eq:UnitarityConstraint}
{\sigma_\mathrm{ann}}_{\;\stau^+\,\stau^-\rightarrow
 h\,h}
 ~\leq ~\sigma_{\mathrm{unit},s}~=~\frac{4 \pi}{|\vec{p}_\mathrm{in}|^2}\;.
\end{equation}
If the bound is not respected, this signals that the perturbative calculation of the cross section 
is no longer valid. In our case, the unitarity bound
translates into a constraint  on the $\mu$ parameter. The bound is practically
independent of $m_h$, and reads
\begin{equation}\label{eq:UnitarityBound2}
 \mu_\mathrm{eff}\,\frac{\tan\beta}{50}
 ~\lesssim~4.1\,\mathrm{TeV}\times\left(\frac{m_{\widetilde{\tau}_1}}{150\gev}\right)\;.
\end{equation}
We find that the annihilation channel into Higgs pairs relative to the
annihilation into gauge bosons can well be larger by factors of
$\mathcal{O}(10^3)$ without violating the unitarity 
bound~\eqref{eq:UnitarityBound2}. 

\section{Primordial staus}
\label{primordial}

\subsection{BBN constraints}

Various cosmological constraints on the stau yield $Y_\stau \equiv  Y_{\stau^+}
+Y_{\stau^-}$ have been explored in the literature.  In a scenario where the LSP
is very weakly coupled, the NLSP decays a considerable time after the start of
BBN. The decay products of such long-lived particles can alter the primordial
light element abundances. This leads to constraints on the released
electromagnetic and hadronic energy 
\cite{Cyburt:2002uv,Kawasaki:2004qu,Jedamzik:2006xz}.  The constraints on the
stau relic abundance from decays depend on the stau lifetime and mass as well
as on the electromagnetic and hadronic branching ratios. For reasonable values
of the stau mass, the hadronic branching fraction is typically  $\lesssim
\mathcal{O}(10^{-3})$ which leads to rather weak constraints. Stronger
constraints come from electromagnetic energy injection, especially at late
times. Here the bounds can be as strong as $m_\stau Y_\stau \lesssim 10^{-13} \gev$
\cite{Kawasaki:2007xb}. However, typically the stau NLSP abundance and lifetime
can satisfy the limits obtained from BBN by considering NLSP decays alone
\cite{Fujii:2003nr,Feng:2004mt,Ellis:2003dn,Cerdeno:2005eu,Jedamzik:2005dh,%
Steffen:2006hw,Buchmuller:2006nx,Cyburt:2006uv,Pradler:2006hh}

In addition to injecting energetic showers into the plasma through decays,
negatively charged particles can form bound  states with light nuclei, which can
lead to a drastic overproduction of $^6$Li \cite{Pospelov:2006sc} and $^9$Be
\cite{Pospelov:2008ta}.  This leads to strong constraints on the stau yield
$Y_{\stau^-}$ for lifetimes longer than a few thousand seconds. While
\cite{Jedamzik:2007qk} gives a rather conservative upper bound of $Y_{\stau^-}
\lesssim 10^{-14}$ derived from $^6$Li alone, \cite{Pospelov:2008ta}
takes into account $^6$Li as well as $^9$Be leading to $Y_{\stau^-} \lesssim
10^{-15}$ which translates into $Y_{\stau} \lesssim 2 \cdot 10^{-15}$ in the
absence of a stau anti-stau asymmetry.

\subsection{Stau relic abundance}

The relic abundance of a (meta)stable particle can be calculated using the
Boltzmann equation. For the stau yield $Y_\stau$,  (again in the absence of a
stau anti-stau asymmetry) it takes the form \cite{Gondolo:1990dk}
\begin{equation}\label{eq:Boltzmannyield}
 \frac{\D Y_{\stau}}{\D x}
 ~=~-\sqrt{\frac{2 g_*}{45}}\,\pi\, M_\mathrm{P}\, 
 \frac{m_{\stau_1}}{x^2}\,\left\langle\sigma_\mathrm{ann}\, v\right\rangle\, 
 \left( Y^2_{\sltau} -
 Y^2_{\sltau,\mathrm{eq}}\right) \;,
\end{equation}
where $M_\mathrm{P}=2.43\cdot 10^{18} \gev$ is the reduced Planck mass,
$Y^2_{\sltau,\mathrm{eq}}$ is the abundance in thermal equilibrium  and $g_*
\simeq 85$ represents the effective number of degrees of freedom at  the stau
freeze out (cf.~\cite{Gondolo:1990dk}).

It is well known that the relic abundance of a (meta)stable particle is
inversely proportional to its annihilation cross section times its mass (see
e.g.\ \cite{Drees:2004jm}). In the case where the stau freeze-out is  dominated
by the Higgs channel, we can write the solution to~\eqref{eq:Boltzmannyield} as
\begin{equation}\label{StauYieldApproximation2}
 Y_\stau ~=~10^{-15} \left( \frac{10^{-5} \gev^{-2}}{\langle\sigma v\rangle }\right) 
\left( \frac{200 \gev}{m_{\stau_1}}\right) \;.
\end{equation}
If stau and Higgs are not mass degenerate ($m_{\stau_1} - m_h \gtrsim 5 \gev$),
the annihilation cross section~\eqref{eq:ThermalCrossSection} is practically
independent of $m_h$ and depends only on the stau  mass $m_{\stau_1}$ and the
stau-Higgs coupling $\propto \mu\,\tan\beta\,
\sin{2\theta_\stau}=\mu_\mathrm{eff}\,\tan\beta$.  Our result for
the  yield can then be written as
\begin{equation}\label{eq:StauYieldApproximation}
 Y_\stau =1.4\cdot10^{-15} \left( \frac{m_{\stau_1}}{150\gev}\right)^5 \left( \frac{1 \tev}{\mu}\right)^4 
\left( \frac{50}{\tan\beta}\right)^4 \left( \frac{1}{\sin{2\theta_\stau}}\right)^4 
\;.
\end{equation}
Here we neglected subleading effects of the order $10\,\%$ like e.g.~Sommerfeld
enhancement.\footnote{See e.g.~\cite{Berger:2008ti} for an explanation of the  Sommerfeld
effect and an estimate of the errors in the general case of a charged relic.}
Combining our result~\eqref{eq:StauYieldApproximation} with the unitarity
bound~\eqref{eq:UnitarityBound2} we find that the lowest allowed abundance is
$Y^\text{min}\sim 4\cdot 10^{-18}$ for $m_{\stau_1}=120\gev$.
As explained in Section~\ref{sec:muconstraints}, the minimal, theoretically
viable abundance might well turn out to be larger than this value. From the conservative
bound \eqref{eq:mubound2}
we infer however that nevertheless $Y^\text{min} \lesssim 10^{-15}$.

\subsection{A comment on stau asymmetries}

\begin{figure}[h]
\centerline{\subfigure[]{\CenterEps{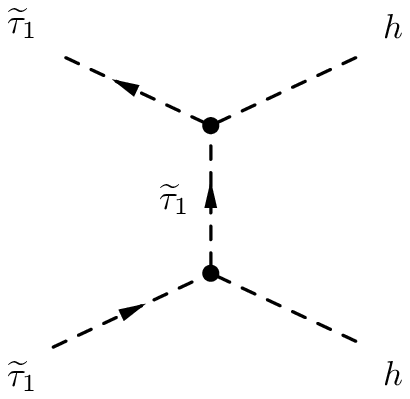}}\quad\quad\subfigure[]{\CenterEps{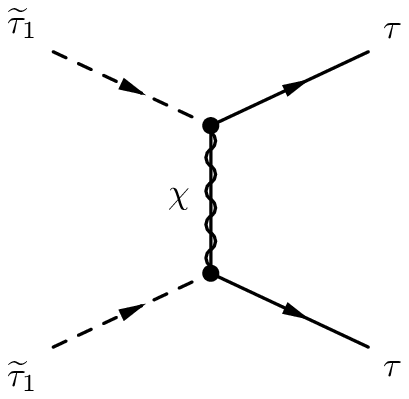}}}
\caption{Examples for slepton number conserving (a) and violating (b) annihilation processes.}
\label{fig:SleptonNumberNonConservation}
\end{figure}

So far we have assumed that there is no asymmetry in the stau sector, that is,
there are as many $\widetilde{\tau}^+$ as $\widetilde{\tau}^-$ degrees of
freedom. 
On the other hand, a large class of baryogenesis mechanisms rely on
$(B+L)$ violation by sphalerons \cite{Klinkhamer:1984di,Kuzmin:1985mm}, which
leads to an excess of baryons over anti-baryons if there are more anti-leptons
than leptons. In particular, leptogenesis \cite{Fukugita:1986hr} falls into this
class. From this point of view it appears natural to assume that there is also
an asymmetry in the slepton sector at the time of stau annihilation and freeze
out. 
Now it is important to distinguish between slepton number conserving and slepton
number violating annihilation processes
(Figure~\ref{fig:SleptonNumberNonConservation}).

As we have seen, in settings with substantial left-right mixing in the stau
sector, the slepton number conserving processes dominate over the violating
ones. On the other hand, the slepton number violating processes are still
effective until the stau relic abundance has reached a value
$Y_{\stau}\sim10^{-12}\dots10^{-13}$ \cite{Asaka:2000zh}. It is then clear that
if the slepton asymmetry in the stau sector is of the order of the baryon
asymmetry, $\eta_b\sim 10^{-10}$, asymmetries will play no role. However, if
there are order one asymmetries in the (s)tau sector, this might have important
implications for the relic abundance of $\stau^-$: in a situation in which there
is a large excess of $\stau^+$ over $\stau^-$, the remaining $\stau^-$ are more
likely to find an annihilation partner, and hence their relic abundance can get
suppressed more strongly. For stau lifetimes where electromagnetic bounds are
not overly restrictive, $\tau_\stau \lesssim 10^7\,\mathrm{s}$, a relatively
large abundance of $\stau^+$,  $Y_{\stau^+}\lesssim10^{-13}$, can still be
consistent with BBN because they do not form bound states with nuclei.

\section{Scenarios with low $\boldsymbol{Y_{\stau}}$}
\label{sec:scenarios}

In this section we present some `top-down motivated settings'  in which the
previously discussed strong suppression of the stau relic abundance occurs. If a
scenario can  have a large stau annihilation cross section is fully determined
by the stau spectrum. The necessary conditions are:
\begin{description} 
\item[Condition 1:] Substantial left-right mixing of the stau mass eigenstates $\stau_1$ and $\stau_2$ driven by a 
large off-diagonal stau mass matrix element $m_{\stau_{\mathrm{LR}}}^2\simeq
-m_\tau\,\mu\,\tan\beta$. This requires large $\mu$ and $\tan\beta$.
\item[Condition 2:] Moderate $m_{\stau_1}$, preferably $m_{\stau_1} \lesssim
200\gev$.
\end{description}
We will check these conditions by looking at the entries of the stau mass
matrix  $m_{\stau_{\mathrm{R}}}^2$, $m_{\stau_{\mathrm{L}}}^2$
and $m_{\stau_{\mathrm{LR}}}^2$  as well as on the mass eigenvalues
$m_{\stau_1}$ and $m_{\stau_2}$. 

It is clear that we could simply write down the desired stau mass matrices. The
purpose of this section however is to present soft mass patterns defined at the
unification scale $M_\mathrm{GUT}$ that lead to mass matrices with the above
properties. To this end, it is useful to recall the (one-loop) renormalization
group equations (RGEs) for the stau soft masses (see e.g.\ \cite{Drees:2004jm})
\begin{subequations}\label{eq:stauRGEs}
\begin{eqnarray}
 8\pi^2\,\frac{\D m_{\stau_{\mathrm{L\,\text{soft}}}}^2}{\D t}
 & = & 
 y_\tau^2\,S_\tau-3g_2^2\,|M_2|^2-g_Y^2\,|M_1|^2-\frac{1}{2}g_Y^2\,S_Y\;,\\
 8\pi^2\,\frac{\D m_{\stau_{\mathrm{R\,\text{soft}}}}^2}{\D t}
 & = & 
 2\,y_\tau^2\,S_\tau-4\,g_Y^2\,|M_1|^2+g_Y^2\,S_Y\;,\label{eq:stauRightRGE}
\end{eqnarray}
\end{subequations}
where 
\begin{subequations}
\begin{eqnarray}
 S_\tau & = & 
 m_{H_1}^2+m_{\stau_{\mathrm{L\,\text{soft}}}}^2+m_{\stau_{\mathrm{R\,\text{soft}}}}^2+|A_\tau|^2
 \;,\label{eq:STau}\\
 S_Y & = & \frac{1}{2}\sum_i Y_i\,m_i^2
\end{eqnarray}
\end{subequations}
with $Y_i$ denoting the hypercharge of the scalar $i$. Note that
$m_{\stau_{\mathrm{L}}}^2 \simeq m_{\stau_{\mathrm{L\,\text{soft}}}}^2$ and
$m_{\stau_{\mathrm{R}}}^2 \simeq m_{\stau_{\mathrm{R\,\text{soft}}}}^2$ as other
contributions are tiny (cf.~\eqref{eq:StauMassMatrix}).

In order to obtain large mixing (condition 1) we have to demand that the right-hand sides (rhs) 
of \eqref{eq:stauRGEs} be similar, assuming coincident stau masses at
the high scale. In addition, we need a large off-diagonal stau mass.
$|\mu|$ which determines the size of $m_{\stau_{\mathrm{LR}}}^2$ together with $\tan\beta$ is fixed at 
the weak scale by the condition of correct electroweak symmetry
breaking which reads (at tree-level)~\cite{Drees:2004jm}
\begin{equation}\label{eq:MuEWSB}
 |\mu|^2=\frac{m_{H_2}^2 \sin^2\beta-m_{H_1}^2 \cos^2\beta}{\cos{2\beta}}-\frac{M_Z^2}{2}
\stackrel{\text{large }\tan\beta}{\simeq} -m_{H_2}^2\;.
\end{equation}
An unsuppressed $m_{\stau_{\mathrm{LR}}}^2$ can typically be realized for
$\mu\sim 1 - 2\tev$. In principle it is  not difficult to get $\mu$ in this
range, however one should mention here that a relatively large $\mu$ might be 
associated with a significant amount of electroweak fine-tuning as can be seen
from \eqref{eq:MuEWSB}.In addition, very large values for $\mu$ might lead to charge
breaking vacua with unacceptably short lifetimes (cf.\ Section~\ref{sec:muconstraints}).

The second condition is already partially fulfilled if the mixing is sizable.
To further reduce the stau masses, the rhs of \eqref{eq:stauRGEs}
should not be too negative.

In what follows we present three scenarios where the desired stau mass patterns
arise and low relic abundances  through the Higgs channel can be achieved. We
use micrOMEGAs 2.0.7 \cite{Belanger:2001fz,Belanger:2006is} to calculate the
relic abundance of the stau NLSP numerically.  
The superpartner spectrum is determined by SOFTSUSY 2.0.18
\cite{Allanach:2001kg} whereas the Higgs mass is calculated  with the
specialized tool FeynHiggs 2.5.1
\cite{Heinemeyer:1998yj,Heinemeyer:1998np,Degrassi:2002fi,Frank:2006yh}.  For
the top quark pole mass, we use the latest best-fit value of
$172.6\gev$~\cite{Group:2008nq}. Experimental constraints on the parameter space
arise primarily through mass limits.  We employ the LEP Higgs bound $m_h\geq 114
\gev$ and $m_{\stau_1}\geq 100\gev$ \cite{staumass}. 
Theoretical constraints, as discussed in
Section~\ref{sec:muconstraints}, are not shown explicitly.

\subsection{CMSSM with large $\boldsymbol{\tan\beta}$}

\begin{figure}[t!] 
  \centering
  \begin{minipage}[b]{7.2 cm}
    \includegraphics[width=7.3cm,height=8cm]{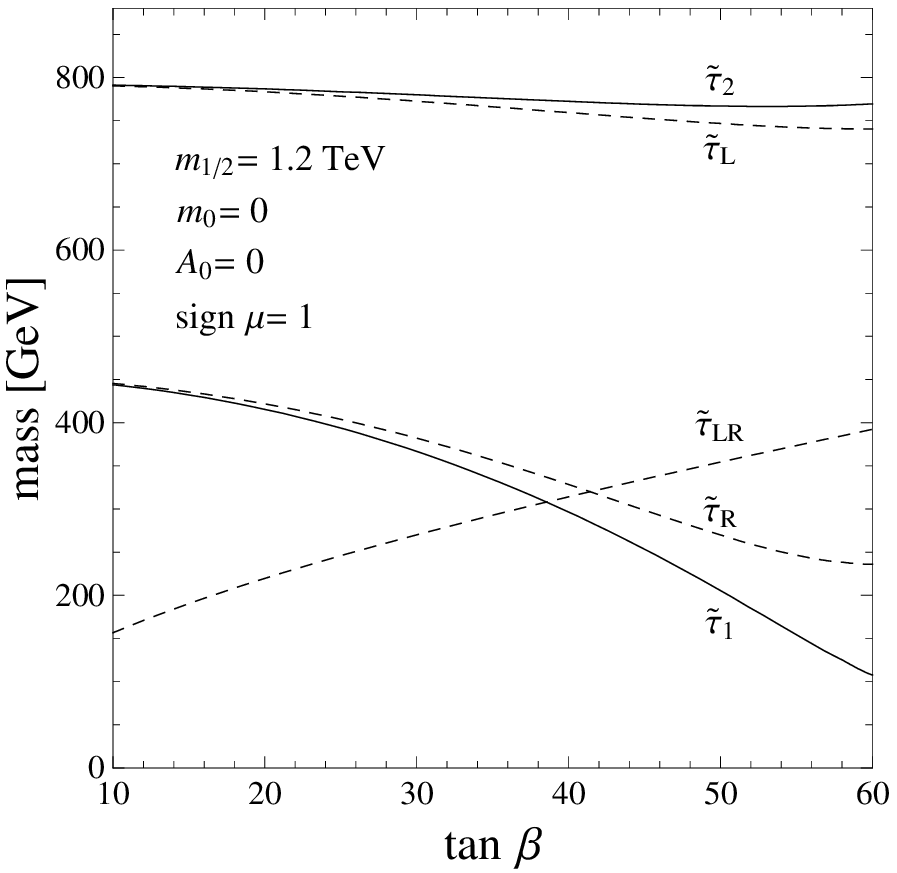}  
  \end{minipage}\hspace*{1cm}
  \begin{minipage}[b]{7.2 cm}
    \includegraphics[width=7.3cm,height=8cm]{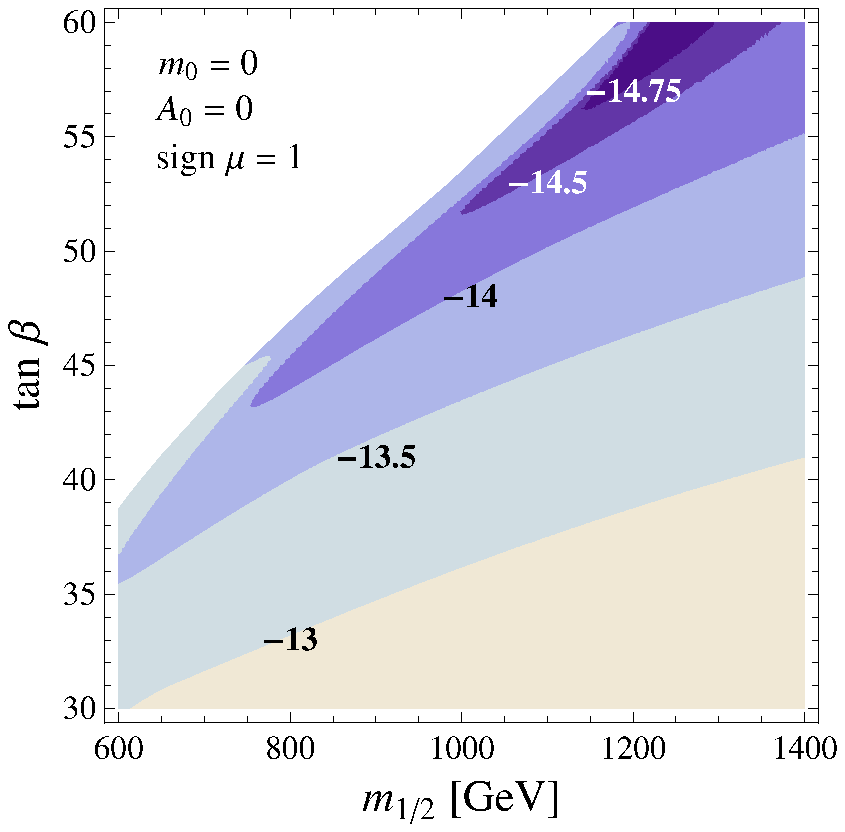}  
  \end{minipage}
  \caption{ Stau relic abundance in the CMSSM. The left panel shows the dependence of the left- and right-handed 
            stau masses, the off-diagonal mass 
            and the mass eigenvalues on $\tan\beta$. In the right panel we plot the logarithm of the stau relic abundance 
            $\log_{10}(Y_\stau)$ in the ($m_{1/2}-\tan\beta$) --plane.
            We find the minimal yield to be around $Y_{\sltau}^\text{min}\simeq 10^{-15}$. 
			The white region in the right panel is excluded due to direct searches.
}
  \label{fig:msugrarelic}
\end{figure}

Let us start with the constrained supersymmetric standard model (CMSSM), which
is defined through  its free parameters $m_{1/2}$, $m_0$, $A_0$, $\tan\beta$ and
$\text{sign}\, \mu$. Although we will see that the annihilation can be more
efficient in other scenarios, we find it nevertheless worthwhile to point out
that also in this scheme a major suppression of the relic abundance is possible.
The important quantity here is $\tan\beta$ which we plot against the stau
spectrum in the left panel of Figure~\ref{fig:msugrarelic}  for a typical stau
NLSP parameter point.

The plot shows that $m_{\stau_{\mathrm{R}}}$ and $m_{\stau_1}$ decrease strongly
with $\tan\beta$ through the tau Yukawa term in the RGE of
$m_{\stau_{\mathrm{R\,\text{soft}}}}^2$, because $y_\tau \propto \tan\beta$.  Since
$\mu$ is practically  independent of $\tan\beta$ the off-diagonal 
mass\footnote{We use the definition
$m_{\stau_{\mathrm{LR}}}=\sqrt{|m_{\stau_{\mathrm{LR}}}^2|}$ in the following.} 
$m_{\stau_{\mathrm{LR}}}$ grows like $\sqrt{\tan\beta}$ which leads to a further
reduction of $m_{\stau_1}$ through left right-mixing.  However, in spite of a
strong off-diagonal stau mass, mixing cannot get close to maximal, because of a
large difference $m_{\stau_{\mathrm{L}}}^2-m_{\stau_{\mathrm{R}}}^2$. 
We conclude that condition 2 can easily be satisfied while condition 1 only
partially. In summary, we obtain a significant enhancement of the
stau annihilation cross section through the Higgs channel at large $\tan\beta$ 
which is however limited by the mixing angle. To illustrate the effect,
Figure~\ref{fig:msugrarelic} shows the stau relic abundance in the CMSSM as a function of
$\tan\beta$.

\subsection{Non-universal Higgs masses (NUHM)}

\begin{figure}[t!]
  \centering
  \begin{minipage}[b]{7.2 cm}
    \includegraphics[width=7.3cm,height=8cm]{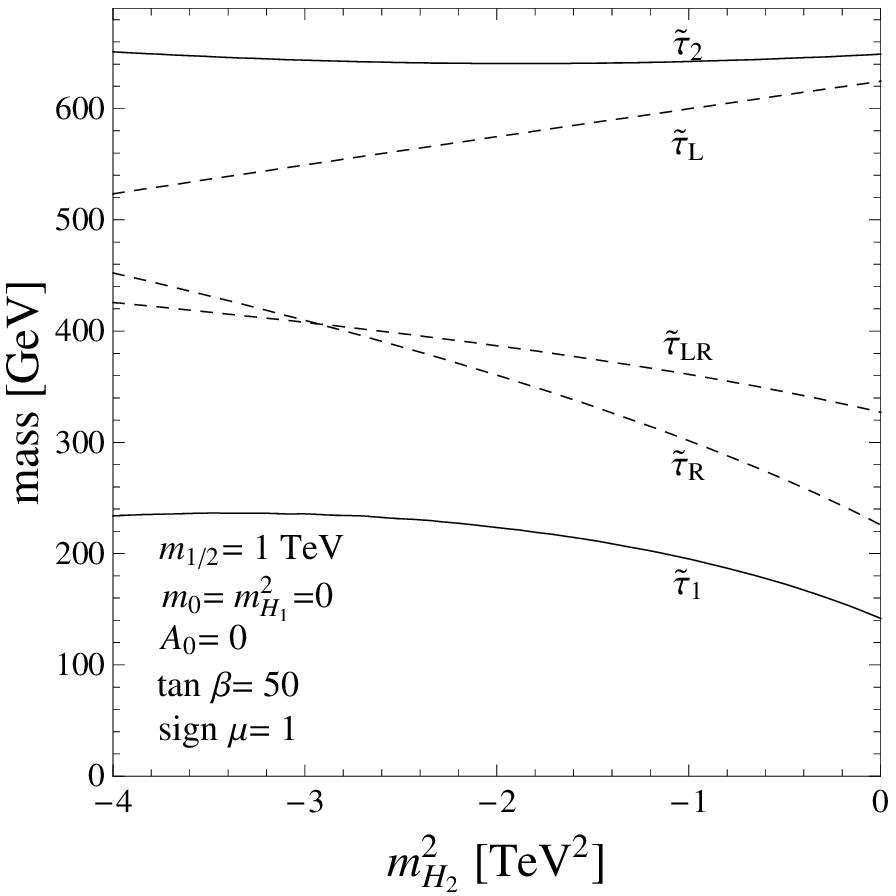}  
  \end{minipage}\hspace*{1cm}
  \begin{minipage}[b]{7.2 cm}
    \includegraphics[width=7.3cm,height=8.02cm]{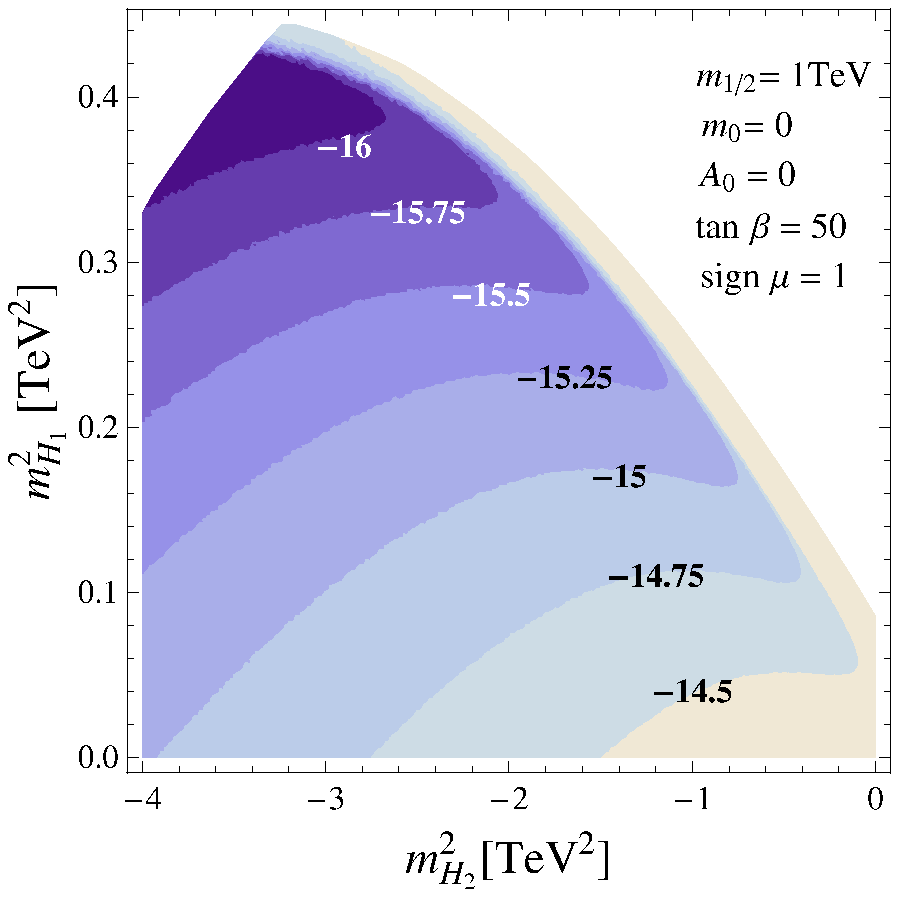}  
  \end{minipage}
  \caption{Stau relic abundance in the NUHM. The left panel shows the dependence of the left- and right-handed 
            stau masses, the off-diagonal mass 
            and the mass eigenvalues on the soft Higgs mass $m_{H_2}^2$. 
            In the right panel we plot the logarithm of the stau relic abundance 
            $\log_{10}(Y_\stau)$ for different $m_{H_1}^2$ and $m_{H_2}^2$.
            Some of the low $\stau$ yield regions might be
			excluded by theoretical constraints (cf.\
			Section~\ref{sec:muconstraints}). Again, the white region in the right panel is excluded. }
  \label{fig:NuhmSpectrum}
\end{figure}

In the NUHM we can vary -- in addition to the parameters of the CMSSM -- the
down- and up-type soft Higgs  masses squared at the GUT scale, $m_{H_1}^2$ and
$m_{H_2}^2$. We employ the GUT scale stability constraint $m_{H_{1,2}}^2+|\mu^2|
\geq 0$ to avoid dangerous vacua and  electroweak symmetry breaking at the GUT
scale \cite{Ellis:2002iu}. It is instructive to investigate the additional 
effects on the stau spectrum compared to the CMSSM which arise through the
variation of the soft Higgs masses: increasing $m_{H_1}^2$ leaves
$m_{\stau_{\mathrm{LR}}}$ unchanged, but it reduces $m_{\stau_{\mathrm{L}}}$ and
$m_{\stau_{\mathrm{R}}}$ dominantly through the Yukawa term  in the soft mass
RGEs \eqref{eq:stauRGEs} which contains $m_{H_1}^2$.  However more interesting
for us is the impact of $m_{H_2}^2$ which we show in the left panel of
Figure~\ref{fig:NuhmSpectrum}.
We observe that a growth of $m_{H_2}^2$ reduces  $m_{\stau_{\mathrm{R}}}$ and
enlarges $m_{\stau_{\mathrm{L}}}$. As $m_{H_2}^2$ does not appear in the  Yukawa
term of the RGEs \eqref{eq:stauRGEs}, this effect arises through the $S_Y$-term.
For the off-diagonal stau mass $m_{\stau_{\mathrm{LR}}}$, it is
important to recall that $|\mu^2|\simeq -m_{H_2}^2$ at the weak scale,
cf.~\eqref{eq:MuEWSB}. Therefore, increasing $m_{H_2}^2$ leads to a
suppression of $m_{\stau_{\mathrm{LR}}}^2$.

The low yield parameter region is at relatively large negative $m_{H_2}^2$, where the GUT 
stability constraint can still be satisfied and again at large $\tan\beta$. Here both, the left-right mixing as well as the off-diagonal 
stau mass $m_{\stau_{\mathrm{LR}}}$, can further be enhanced compared to the CMSSM case. Stau masses which are slightly 
larger can be reduced by a positive $m_{H_1}^2$. We conclude that both conditions can be satisfied 
in this region, leading to an extremely suppressed stau abundance.

\subsection{Scenarios with non-universal gaugino masses (NUGM)}

\begin{figure}[t!]
  \centering
  \begin{minipage}[b]{7.3 cm}
    \includegraphics[width=7.2cm,height=8cm]{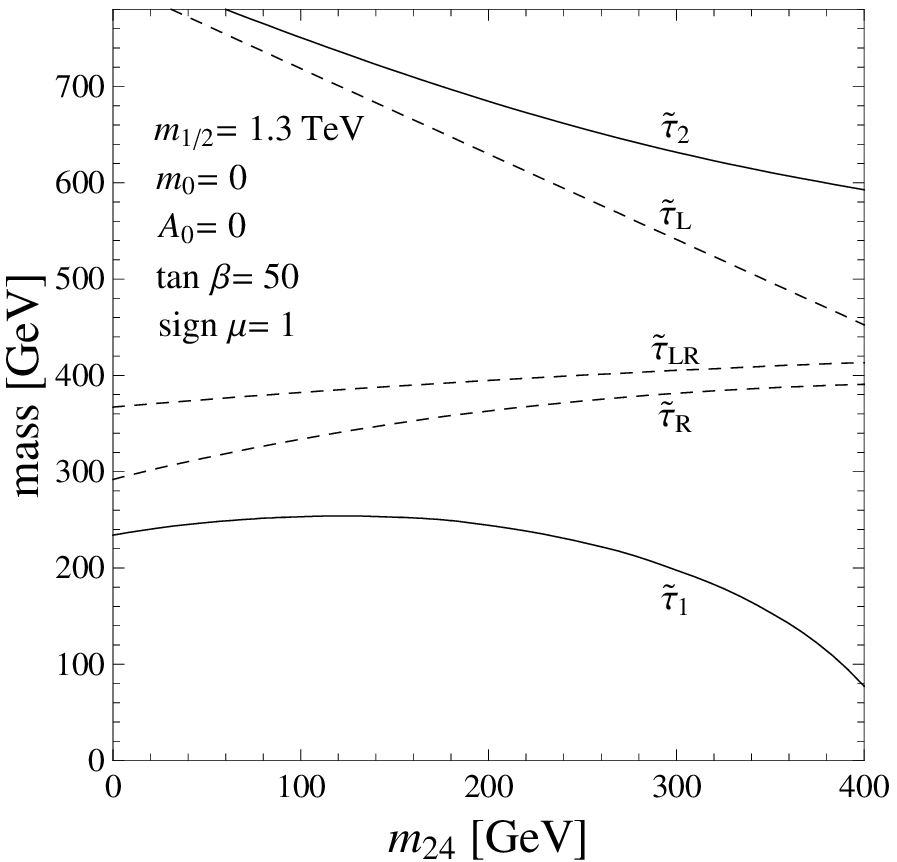}  
  \end{minipage}\hspace*{1cm}
  \begin{minipage}[b]{7.3 cm}
    \includegraphics[width=7.2cm,height=8cm]{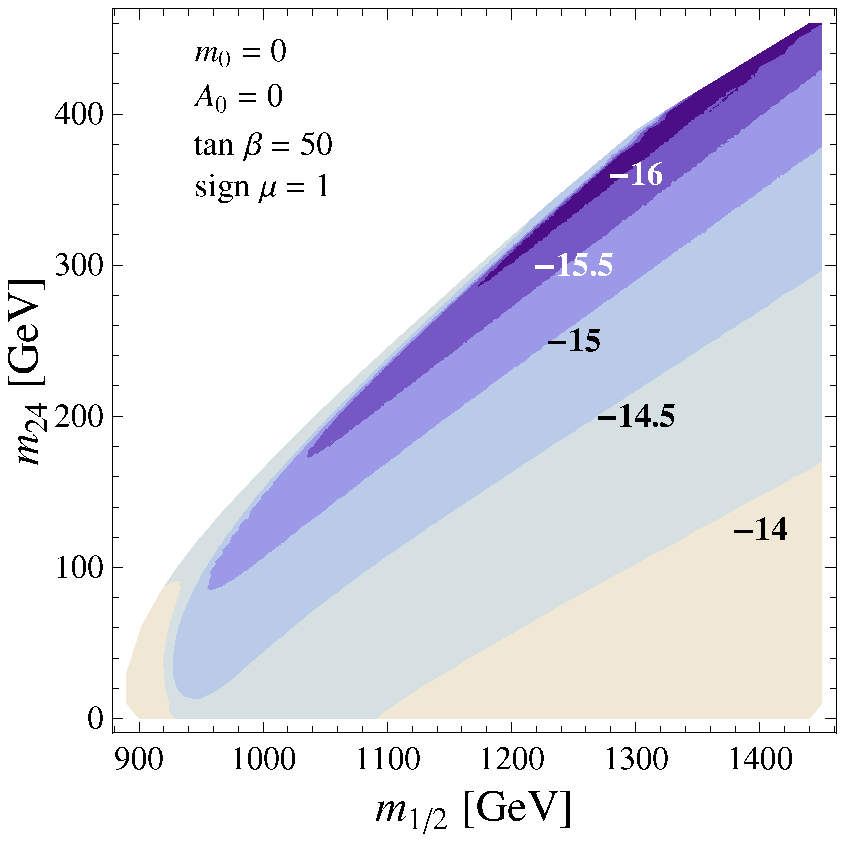}  
  \end{minipage}
  \caption{Stau relic abundance in the NUGM. The left panel shows the dependence of the left- and right-handed 
            stau masses, the off-diagonal mass 
            and the mass eigenvalues on $m_{24}$. In the right panel we plot the logarithm of the stau relic abundance 
            $\log_{10}(Y_\stau)$ in the ($m_{1/2}-m_{24}$) --plane.
                      Some of the low $\stau$ yield regions might be
			excluded by theoretical constraints (cf.\
			Section~\ref{sec:muconstraints}).
	    Once more, the white region in the right panel is excluded. 
}
  \label{fig:Rossrelic}
\end{figure}
The possibility of having non-universal gaugino masses as high scale boundary
conditions even for unified theories has been realized long ago
\cite{Ellis:1985jn}.  For concreteness we focus here on SU(5) GUTs and assume
that supersymmetry be broken by $F$-term vacuum expectation values of chiral
fields. These fields have to  transform as the symmetric product of two adjoint
representations of the GUT group, but not necessarily as  singlets. In the
following we concentrate on the two smallest possible representations for the
supersymmetry  breaking fields, which are simply the singlet and the
$\boldsymbol{24}$plet.

The high-scale mass patterns of the gauginos of SU(3)$_C$, SU(2)$_\mathrm{L}$
and U(1)$_Y$ turn out to be given as linear combinations of singlet ($m_{1/2}$)
and $\boldsymbol{24}$plet ($m_{24}$) contributions \cite{Huitu:1999ac},
\begin{eqnarray}
 M_1&=&m_{1/2}-0.5\,m_{24} \;,\nonumber\\
 M_2&=&m_{1/2}-1.5\,m_{24} \;,\nonumber\\
 M_3&=&m_{1/2}+m_{24}      \;.\label{eq:GUTGauginoMasses124}
\end{eqnarray}
The only new parameter compared to the CMSSM is the mass arising from the
$\boldsymbol{24}$plet, $m_{24}$. Its impact on the stau spectrum is illustrated
in the left panel of Figure~\ref{fig:Rossrelic}.

$m_{\stau_{\mathrm{L}}}$ decreases quickly for growing $m_{24}$ as the smaller
$M_2$ reduces the SU(2)$_\mathrm{L}$ gauge contribution to
$m_{\stau_{\mathrm{L\,\text{soft}}}}^2$. $m_{\stau_{\mathrm{R}}}$ increases
slightly, although the U(1)$_Y$ gauge contribution to
$m_{\stau_{\mathrm{R\,\text{soft}}}}^2$ shrinks. This is because the reduction
of $m_{\stau_{\mathrm{R\,\text{soft}}}}^2$ through the Yukawa term in
\eqref{eq:stauRightRGE} is less effective for smaller
$m_{\stau_{\mathrm{L\,\text{soft}}}}^2$. The off-diagonal mass matrix entry
$m_{\stau_{\mathrm{LR}}}^2 \propto \mu$ grows slowly as $\mu$ gets a contribution
from the increasing gluino mass $M_3$.

It turns out that diagonal and off-diagonal stau soft masses can come very
close, leading to maximal left-right mixing in the stau
sector with a strong reduction of $m_{\sltau[1]}$ driven by a large
$m_{\stau_{\mathrm{LR}}}$. As this is the key ingredient for strongly enhanced
couplings of $\sltau[1]$ to the light Higgs, the stau relic abundance becomes very small, as can be seen in 
the right panel of Figure~\ref{fig:Rossrelic}.

\subsection{Further remarks}

Let us briefly summarize the main results of this section. We have seen that a
strong suppression of the  stau relic abundance can be achieved such that all
cosmological constraints can be evaded. We also checked that the shown regions
of parameter space are not in conflict with  precision measurements of $b
\rightarrow s\,\gamma$ and the Tevatron limit of $B_s \rightarrow \mu^+\mu^-$.
Furthermore, the parameter regions with low relic abundance  are all within the
$2\,\sigma$ interval of the measured anomalous magnetic moment of the muon, the
only exception  being the CMSSM for $m_{1/2}\gtrsim 1100 \gev$.
Comparing the low yield
regions with the CBBN bound $Y_\stau \lesssim 2\cdot 10^{-15}$ we see that 
even in the CMSSM there remains some viable region in parameter space while for
the NUHM and NUGM the stau relic abundance can be even
smaller. Further analysis of the experimental signatures of such low stau yield
regions appears desirable.

\section{Prospects for the LHC}
\label{sec:Prospects}

A very appealing property of our scenario is that it will be tested at the LHC.
Given the stau spectrum, the stau-Higgs coupling is fully determined. It is then
easy to see the impact of the Higgs channel for stau annihilation in the early
universe.
 
A common feature of the low yield regions is a rather small stau mass,
$m_{\stau_1} \lesssim 250\gev$, while the other SUSY particles may be quite
heavy. In general, the prospects for the LHC depend mainly on the mass scale of
the colored superpartners. If they are not too heavy, as can be the case e.g.~in
the NUHM, they will be produced  in large numbers at the LHC due to their large
cross sections. In this case one has a good chance to measure  the whole SUSY
spectrum in the cascade decays of gluinos and squarks and it should not be too
difficult to extract information about the stau-Higgs coupling.

If, however, the mass scale of colored particles exceeds $2-2.5\tev$, their
production will  become very rare or even impossible at the LHC
\cite{Baer:2003wx}.  But even in this case, being rather light, staus could
still be pair produced, e.g.~in the Drell-Yan process $q \, \bar{q} \rightarrow
\stau_1^- \,\stau_1^+ $ through a virtual photon or $Z$ boson
\cite{Eichten:1984eu}.  The number of produced stau pairs for different stau
masses are shown in Table~\ref{drellyan}.   Further details on the stau spectrum might be extracted from the
Drell-Yan production of  $\stau_1 \, \snu_\tau$, $\stau_1 \, \stau_2$ and
$\stau_2 \, \stau_2$, if kinematically accessible.

\begin{table}[!ht!]
  \centering
  \renewcommand{\arraystretch}{1.7}
  \begin{tabular}{||c||c|c|c|c||}
    \hline
Mass & $150\gev$ & $200\gev$ & $300\gev$ & $400\gev$  
\\      \hline
$\#_{\stau_\text{R}}$  & $2000$  & $800$ & $200$ & $60$  
\\      \hline
$\#_{\stau_\text{L}}$  & $6000$  & $2000$ & $450$ & $150$  
\\ \hline
  \end{tabular}
  \caption{Estimated number of produced stau pairs in Drell-Yan processes at the
		   LHC for integrated luminosity of  $100 \, \text{fb}^{-1}$ extracted
		   from Figure~1 of \cite{Bozzi:2004qq}. The number of produced mass
		   eigenstates depends on the mixing angle and should lie in between the values given for left- and
		   right-handed staus.}
\label{drellyan}
\end{table}
Another interesting pair production mechanism in our scenario is gluon-gluon fusion.
Here two gluons generate a fermion loop (preferably a top-loop) to which a virtual Higgs boson is attached 
which finally decays into a stau pair. 
In an early study \cite{delAguila:1990yw} the cross section for this process was found to be three orders of magnitude
below the corresponding Drell-Yan cross section. 
Note that in our scenario, due to the strong stau-Higgs coupling, this
suppression can at least partially be compensated  such that gluon-gluon fusion could be comparable to
the Drell-Yan process.

Altogether we see that even in scenarios in which some superpartners are beyond
the reach of the LHC, one may nevertheless establish the existence of
supergravity in nature along the lines of \cite{Buchmuller:2004rq,Buchmuller:2004tm,Feng:2004mt,Hamaguchi:2004df,Feng:2004yi,Brandenburg:2005he,Martyn:2006as,%
Ellis:2006vu,Kitano:2008sa}.

\section{Conclusions}

We have analyzed stau NLSP scenarios. In contrast to previous studies, we have
not assumed that the stau mass eigenstates be purely right- or left-handed, but
have allowed for non-trivial left-right mixing. In the case of substantial
mixing, the annihilation into Higgs bosons can dominate over other channels,
such that the thermal relic stau abundance, i.e.\ the abundance obtained without
invoking late-time entropy production, can be strongly reduced. This makes it
possible to evade all BBN constraints.  The emerging scenarios have the
advantage that they allow for rather large reheating temperatures, as required
for instance by leptogenesis, and the cold dark matter can be explained in terms of
`thermally' produced gravitinos. Most importantly, all ingredients of our low
stau yield scenarios will be tested at the LHC.

\section*{Acknowledgments}
We would like to thank Alejandro Ibarra and David Straub for useful discussions.
We are indebted to the referee for important comments on an earlier version of
this paper.
This work has been supported by the SFB-Transregio 27 ``Neutrinos and  Beyond''
and by the DFG cluster of excellence ``Origin and  Structure of the Universe''.

\appendix
\section{Stau Masses \& Mixings}
\label{app:StauMassMatrix}

Let us briefly introduce our conventions concerning the masses and mixings of the stau.
We assume that there is no mixing between different slepton generations and take $\mu$ and $A^\tau$ to be real parameters, 
such the stau mass matrix can be written as
\begin{align}
\hspace{-3mm}
\mathcal{M}_{\sltau}^2&=\begin{pmatrix} 
m_{\stau_{\mathrm{L\,\text{soft}}}}^2 + (\sin^2{\theta_\mathrm{W}}-\frac{1}{2}) M_Z^2 \cos{2\beta}+m_\tau^2 & -m_\tau (A^{\tau} 
+ \mu \tan\beta) \nonumber\\
-m_\tau (A^{\tau} + \mu \tan\beta)  &  m_{\stau_{\mathrm{R\,\text{soft}}}}^2 -\sin^2{\theta_\mathrm{W}} M_Z^2 \cos{2\beta} +m_\tau^2 
\end{pmatrix}\qquad \label{eq:StauMassMatrix}
\\
&\equiv\begin{pmatrix} 
m_{\stau_{\mathrm{L}}}^2 & m_{\stau_{\mathrm{LR}}}^2\\
m_{\stau_{\mathrm{LR}}}^2 & m_{\stau_{\mathrm{R}}}^2 
\end{pmatrix}\;.
\end{align}
A non-zero off-diagonal element $m_{\stau_{\mathrm{LR}}}^2$ leads to a mixing of the chiral states $\stau_\text{L}$ and $\stau_\text{R}$. 
We can diagonalize the stau mass matrix by an orthogonal transformation
\begin{equation}
\mathcal{O^T}\, \mathcal{M}_{\sltau}^2\, \mathcal{O} 
~=~\begin{pmatrix}m_{\sltau[1]}^2 & 0\\0& m_{\sltau[2]}^2\end{pmatrix}\;.
\end{equation}
The mass eigenvalues are
\begin{equation}
m_{\sltau[1,2]}^2~ =~\frac{1}{2} \left[ m_{\stau_{\mathrm{L}}}^2 + m_{\stau_{\mathrm{R}}}^2 \mp 
\sqrt{(m_{\stau_{\mathrm{L}}}^2 - m_{\stau_{\mathrm{R}}}^2)^2 
+ 4 m_{\stau_{\mathrm{LR}}}^4}\:\right] 
\;.
\end{equation}
The orthogonal $2\times2$ matrix $\mathcal{O}$ is parameterized by the
stau left-right mixing angle $\theta_{\sltau}$, which relates the mass
eigenstates and the chiral states,
\begin{equation}\label{eq:StauMixing}
\left(\begin{array}{c}
\sltau[1] \\ \sltau[2]
\end{array}\right)
~=~
\begin{pmatrix} 
\cos{\theta_{\sltau}} & \sin{\theta_{\sltau}} \\ -\sin{\theta_{\sltau}} & \cos{\theta_{\sltau}} 
\end{pmatrix} 
\left(\begin{array}{c}\sltau[\mathrm{L}] \\ \sltau[\mathrm{R}]\end{array}\right)
\;.
\end{equation}
The mixing angle $\theta_{\sltau}$ is given by
\label{eq:StauMixingAngle}
\begin{eqnarray}
\cos{\theta_{\sltau}}
& = &
\frac{-m_{\stau_{\mathrm{LR}}}^2}{\sqrt{(m_{\stau_{\mathrm{L}}}^2 
- m_{\sltau[1]}^2)^2+m_{\stau_{\mathrm{LR}}}^4}} \;.
\end{eqnarray}

\enlargethispage{0.8cm}

\bibliography{GauginoMediation}

\providecommand{\bysame}{\leavevmode\hbox to3em{\hrulefill}\thinspace}
\frenchspacing
\newcommand{\origttfamily}{}
\let\origttfamily=\ttfamily
\renewcommand{\ttfamily}{\origttfamily \hyphenchar\font=`\-}

\begin{thebibliography}{10}

\bibitem{Buchmuller:2004rq}
W.~Buchm{\"u}ller, K.~Hamaguchi, M.~Ratz, and T.~Yanagida, Phys. Lett.
  \textbf{B588} (2004), 90, \texttt{hep-ph/0402179}.

\bibitem{Buchmuller:2004tm}
W.~Buchm{\"u}ller, K.~Hamaguchi, M.~Ratz, and T.~Yanagida,
  \texttt{hep-ph/0403203}.

\bibitem{Feng:2004mt}
J.~L. Feng, S.~Su, and F.~Takayama, Phys. Rev. \textbf{D70} (2004), 075019,
  \texttt{hep-ph/0404231}.

\bibitem{Brandenburg:2005he}
A.~Brandenburg, L.~Covi, K.~Hamaguchi, L.~Roszkowski, and F.~D. Steffen, Phys.
  Lett. \textbf{B617} (2005), 99, \texttt{hep-ph/0501287}.

\bibitem{Hamaguchi:2004df}
K.~Hamaguchi, Y.~Kuno, T.~Nakaya, and M.~M. Nojiri, Phys. Rev. \textbf{D70}
  (2004), 115007, \texttt{hep-ph/0409248}.

\bibitem{Feng:2004yi}
J.~L. Feng and B.~T. Smith, Phys. Rev. \textbf{D71} (2005), 015004,
  \texttt{hep-ph/0409278}.

\bibitem{Martyn:2006as}
H.-U. Martyn, \texttt{hep-ph/0605257}.

\bibitem{Ellis:2006vu}
J.~R. Ellis, A.~R. Raklev, and O.~K. {\O}ye, \texttt{hep-ph/0607261}.

\bibitem{Kitano:2008sa}
R.~Kitano, \texttt{arXiv:0806.1057} [hep-ph].

\bibitem{Kim:1983dt}
J.~E. Kim and H.~P. Nilles, Phys. Lett. \textbf{B138} (1984), 150.

\bibitem{Giudice:1988yz}
G.~F. Giudice and A.~Masiero, Phys. Lett. \textbf{B206} (1988), 480.

\bibitem{Fukugita:1986hr}
M.~Fukugita and T.~Yanagida, Phys. Lett. \textbf{B174} (1986), 45.

\bibitem{Davidson:2002qv}
S.~Davidson and A.~Ibarra, Phys. Lett. \textbf{B535} (2002), 25,
  \texttt{hep-ph/0202239}.

\bibitem{Bolz:2000fu}
M.~Bolz, A.~Brandenburg, and W.~Buchm{\"u}ller, Nucl. Phys. \textbf{B606}
  (2001), 518, \texttt{hep-ph/0012052}.

\bibitem{Pradler:2006hh}
J.~Pradler and F.~D. Steffen, Phys. Lett. \textbf{B648} (2007), 224,
  \texttt{hep-ph/0612291}.

\bibitem{Falomkin:1984eu}
I.~V. Falomkin et~al., Nuovo Cim. \textbf{A79} (1984), 193, [Yad.\ Fiz.\ {\bf
  39} (1984), 990].

\bibitem{Khlopov:1984pf}
M.~Y. Khlopov and A.~D. Linde, Phys. Lett. \textbf{B138} (1984), 265.

\bibitem{Ellis:1984eq}
J.~R. Ellis, J.~E. Kim, and D.~V. Nanopoulos, Phys. Lett. \textbf{B145} (1984),
  181.

\bibitem{Cerdeno:2005eu}
D.~G. Cerde\~{n}o, K.-Y. Choi, K.~Jedamzik, L.~Roszkowski, and R.~Ruiz~de
  Austri, JCAP \textbf{0606} (2006), 005, \texttt{hep-ph/0509275}.

\bibitem{Kanzaki:2006hm}
T.~Kanzaki, M.~Kawasaki, K.~Kohri, and T.~Moroi, Phys. Rev. \textbf{D75}
  (2007), 025011, \texttt{hep-ph/0609246}.

\bibitem{Covi:2007xj}
L.~Covi and S.~Kraml, JHEP \textbf{08} (2007), 015, \texttt{hep-ph/0703130}.

\bibitem{Asaka:2000zh}
T.~Asaka, K.~Hamaguchi, and K.~Suzuki, Phys. Lett. \textbf{B490} (2000), 136,
  \texttt{hep-ph/0005136}.

\bibitem{Fujii:2003nr}
M.~Fujii, M.~Ibe, and T.~Yanagida, Phys. Lett. \textbf{B579} (2004), 6,
  \texttt{hep-ph/0310142}.

\bibitem{Ellis:2003dn}
J.~R. Ellis, K.~A. Olive, Y.~Santoso, and V.~C. Spanos, Phys. Lett.
  \textbf{B588} (2004), 7, \texttt{hep-ph/0312262}.

\bibitem{Jedamzik:2005dh}
K.~Jedamzik, K.-Y. Choi, L.~Roszkowski, and R.~Ruiz~de Austri, JCAP
  \textbf{0607} (2006), 007, \texttt{hep-ph/0512044}.

\bibitem{Steffen:2006hw}
F.~D. Steffen, JCAP \textbf{09} (2006), 001, \texttt{hep-ph/0605306}.

\bibitem{Buchmuller:2006nx}
W.~Buchm{\"u}ller, L.~Covi, J.~Kersten, and K.~Schmidt-Hoberg, JCAP
  \textbf{0611} (2006), 007, \texttt{hep-ph/0609142}.

\bibitem{Cyburt:2006uv}
R.~H. Cyburt, J.~Ellis, B.~D. Fields, K.~A. Olive, and V.~C. Spanos, JCAP
  \textbf{0611} (2006), 014, \texttt{astro-ph/0608562}.

\bibitem{Pospelov:2006sc}
M.~Pospelov, Phys. Rev. Lett. \textbf{98} (2007), 231301,
  \texttt{hep-ph/0605215}.

\bibitem{Buchmuller:2006tt}
W.~Buchm{\"u}ller, K.~Hamaguchi, M.~Ibe, and T.~T. Yanagida, Phys. Lett.
  \textbf{B643} (2006), 124, \texttt{hep-ph/0605164}.

\bibitem{Hamaguchi:2007mp}
K.~Hamaguchi, T.~Hatsuda, M.~Kamimura, Y.~Kino, and T.~T. Yanagida, Phys. Lett.
  \textbf{B650} (2007), 268, \texttt{hep-ph/0702274}.

\bibitem{Takayama:2000uz}
F.~Takayama and M.~Yamaguchi, Phys. Lett. \textbf{B485} (2000), 388,
  \texttt{hep-ph/0005214}.

\bibitem{Buchmuller:2007ui}
W.~Buchm{\"u}ller, L.~Covi, K.~Hamaguchi, A.~Ibarra, and T.~Yanagida, JHEP
  \textbf{03} (2007), 037, \texttt{hep-ph/0702184}.

\bibitem{Kersten:2007ab}
J.~Kersten and K.~Schmidt-Hoberg, JCAP \textbf{0801} (2008), 011,
  \texttt{arXiv:0710.4528} [hep-ph].

\bibitem{Drees:2004jm}
M.~Drees, R.~Godbole, and P.~Roy, Hackensack, USA: World Scientific (2004).

\bibitem{Coleman:1977py}
S.~R. Coleman, Phys. Rev. \textbf{D15} (1977), 2929.

\bibitem{Kusenko:1996jn}
A.~Kusenko, P.~Langacker, and G.~Segre, Phys. Rev. \textbf{D54} (1996), 5824,
  \texttt{hep-ph/9602414}.

\bibitem{Ferreira:2001tk}
P.~M. Ferreira, Phys. Lett. \textbf{B512} (2001), 379, \texttt{hep-ph/0102141}.

\bibitem{Griest:1989wd}
K.~Griest and M.~Kamionkowski, Phys. Rev. Lett. \textbf{64} (1990), 615.

\bibitem{Cyburt:2002uv}
R.~H. Cyburt, J.~R. Ellis, B.~D. Fields, and K.~A. Olive, Phys. Rev.
  \textbf{D67} (2003), 103521, \texttt{astro-ph/0211258}.

\bibitem{Kawasaki:2004qu}
M.~Kawasaki, K.~Kohri, and T.~Moroi, Phys. Rev. \textbf{D71} (2005), 083502,
  \texttt{astro-ph/0408426}.

\bibitem{Jedamzik:2006xz}
K.~Jedamzik, Phys. Rev. \textbf{D74} (2006), 103509, \texttt{hep-ph/0604251}.

\bibitem{Kawasaki:2007xb}
M.~Kawasaki, K.~Kohri, and T.~Moroi, Phys. Lett. \textbf{B649} (2007), 436,
  \texttt{hep-ph/0703122}.

\bibitem{Pospelov:2008ta}
M.~Pospelov, J.~Pradler, and F.~D. Steffen, \texttt{arXiv:0807.4287} [hep-ph].

\bibitem{Jedamzik:2007qk}
K.~Jedamzik, \texttt{arXiv:0710.5153 [hep-ph]}.

\bibitem{Gondolo:1990dk}
P.~Gondolo and G.~Gelmini, Nucl. Phys. \textbf{B360} (1991), 145.

\bibitem{Berger:2008ti}
C.~F. Berger, L.~Covi, S.~Kraml, and F.~Palorini, \texttt{arXiv:0807.0211}
  [hep-ph].

\bibitem{Klinkhamer:1984di}
F.~R. Klinkhamer and N.~S. Manton, Phys. Rev. \textbf{D30} (1984), 2212.

\bibitem{Kuzmin:1985mm}
V.~A. Kuzmin, V.~A. Rubakov, and M.~E. Shaposhnikov, Phys. Lett. \textbf{B155}
  (1985), 36.

\bibitem{Belanger:2001fz}
G.~Belanger, F.~Boudjema, A.~Pukhov, and A.~Semenov, Comput. Phys. Commun.
  \textbf{149} (2002), 103, \texttt{hep-ph/0112278}.

\bibitem{Belanger:2006is}
G.~Belanger, F.~Boudjema, A.~Pukhov, and A.~Semenov, Comput. Phys. Commun.
  \textbf{176} (2007), 367, \texttt{hep-ph/0607059}.

\bibitem{Allanach:2001kg}
B.~C. Allanach, Comput. Phys. Commun. \textbf{143} (2002), 305,
  \texttt{hep-ph/0104145}.

\bibitem{Heinemeyer:1998yj}
S.~Heinemeyer, W.~Hollik, and G.~Weiglein, Comput. Phys. Commun. \textbf{124}
  (2000), 76, \texttt{hep-ph/9812320}.

\bibitem{Heinemeyer:1998np}
S.~Heinemeyer, W.~Hollik, and G.~Weiglein, Eur. Phys. J. \textbf{C9} (1999),
  343, \texttt{hep-ph/9812472}.

\bibitem{Degrassi:2002fi}
G.~Degrassi, S.~Heinemeyer, W.~Hollik, P.~Slavich, and G.~Weiglein, Eur. Phys.
  J. \textbf{C28} (2003), 133, \texttt{hep-ph/0212020}.

\bibitem{Frank:2006yh}
M.~Frank et~al., JHEP \textbf{02} (2007), 047, \texttt{hep-ph/0611326}.

\bibitem{Group:2008nq}
CDF, \texttt{arXiv:0803.1683} [hep-ex].

\bibitem{staumass}
{{LEPSUSYWG}, {ALEPH}, {DELPHI}, {L3}, and {OPAL} experiments}, note
  LEPSUSYWG/02-05.1,
  \texttt{(http://lepsusy.web.cern.ch/lepsusy/Welcome.html)}.

\bibitem{Ellis:2002iu}
J.~R. Ellis, T.~Falk, K.~A. Olive, and Y.~Santoso, Nucl. Phys. \textbf{B652}
  (2003), 259, \texttt{hep-ph/0210205}.

\bibitem{Ellis:1985jn}
J.~R. Ellis, K.~Enqvist, D.~V. Nanopoulos, and K.~Tamvakis, Phys. Lett.
  \textbf{B155} (1985), 381.

\bibitem{Huitu:1999ac}
K.~Huitu, T.~Kobayashi, K.~Puolamaki, and Y.~Kawamura, Prepared for EPS-HEP 99,
  Tampere, Finland, 15-21 Jul 1999.

\bibitem{Baer:2003wx}
H.~Baer, C.~Balazs, A.~Belyaev, T.~Krupovnickas, and X.~Tata, JHEP \textbf{06}
  (2003), 054, \texttt{hep-ph/0304303}.

\bibitem{Eichten:1984eu}
E.~Eichten, I.~Hinchliffe, K.~D. Lane, and C.~Quigg, Rev. Mod. Phys.
  \textbf{56} (1984), 579.

\bibitem{Bozzi:2004qq}
G.~Bozzi, B.~Fuks, and M.~Klasen, Phys. Lett. \textbf{B609} (2005), 339,
  \texttt{hep-ph/0411318}.

\bibitem{delAguila:1990yw}
F.~del Aguila and L.~Ametller, Phys. Lett. \textbf{B261} (1991), 326.

\end{thebibliography}
\bibliographystyle{NewArXiv}

\end{document}